# Viscosity of a multispecies plasma containing hydrogen and high-Z ions.

Mikhail Dorf

*Lawrence Livermore National Laboratory, Livermore, California, 94550, USA*

Materials containing hydrogen and high-Z ion species are used in a variety of high energy density physics (HED) applications, and it becomes increasingly important to investigate transport properties of these substances as they become fully-ionized multispecies plasmas. In particular, of significant practical interest, is the analysis of the plasma viscosity as it can determine stability and turbulence properties of the HED system. It has been pointed out by D. Ryutov [Phys. Plasmas **3**, 4336 (1996)] that the presence of a substantial amount of hydrogen in a multispecies plasma can significantly enhance its viscosity. The present work reports on a detailed analysis of this effect and the accurate calculation of the corresponding viscosity coefficient.

*Introduction.*–Materials containing a considerable fraction of hydrogen or its isotopes (e.g., $CH_2$, $CD_2$, LiH) are utilized in a great variety of high energy density physics and inertial confinement fusion (ICF) applications. For instance, polystyrene (CH) or polyimide ($C_{22}H_{10}N_2O_4$) ablator shells are employed in the design of ICF capsules [1-2]. Deuterated polystyrene (CD) fibers have been considered for use as liners for wire array Z-pinches [3]. Local spherical blankets composed of the lithium-hydride (LiH) material can be potentially employed in ICF pulsed systems for tritium breeding and efficient (MHD) energy conversion [4-5]. In addition, $CH_2$ plastic disk targets are extensively used in HED laboratory astrophysics experiments for laser-produced fast plasma flows [6-8]. It is therefore of significant practical interest to investigate the hydrodynamic transport properties of these materials as they become fully-ionized multispecies plasmas.

The present work reports on the calculation of the viscosity coefficient, $\eta$, which affects hydrodynamic stability and determines turbulence properties of the aforementioned HED systems. It has been previously pointed out by D. Ryutov [9] that the plasma viscosity can become enhanced by an order of magnitude due to the presence of hydrogen or a hydrogen isotope. Here, a quantitative analysis of this effect is performed, and an accurate expression for the multispecies plasma viscosity is derived. While the analysis is straightforward and is based on a well-known formalism [10-13], we are not aware of such calculation being done, and believe that the results of this work may prove useful in detailed studies of multispecies HED plasmas.

*Qualitative analysis.*–Consistent with the practical applications mentioned earlier, we consider highly-collisional dense plasmas where the mean free path of a hydrogen species, $\lambda_{HZ}$ is small compared to the characteristic length scale for variations in the species flow velocities, $L$,

$$\lambda_{HZ}/L \ll 1. \qquad (1)$$

In such highly-collisional multispecies plasmas, the hydrogen species (H) equilibrates its flow velocity to that of the heavier high-Z ions (Z) over a short collisional time scale, $\tau_c \sim \lambda_{HZ}/V_{TH}$, where $V_{TH}$ is the hydrogen thermal velocity. Considering here only frictional and viscous forces, the momentum fluid equations describing much slower relaxation of the common shear flow velocity $\mathbf{V} = V_x(y)\hat{\mathbf{x}}$ are then given by

$$\frac{\partial(M_H n_H V_x)}{\partial t} = -\frac{V_{TH}}{\lambda_{HZ}} M_H n_H \Delta V_x + \frac{\partial}{\partial y}\eta_H \frac{\partial V_x}{\partial y}, \qquad (2)$$

$$\frac{\partial(M_Z n_Z V_x)}{\partial t} = \frac{V_{TZ}}{\lambda_{ZH}} M_Z n_Z \Delta V_x + \frac{\partial}{\partial y}\eta_Z \frac{\partial V_x}{\partial y}. \qquad (3)$$

Here, $M$, $V_T$, $n$, $\lambda$, and $\eta$ are the species mass, thermal velocity, number density, collisional mean free path, and viscosity coefficient, respectively, and $\Delta V_x$ is the small difference between the hydrogen and high-Z species flow velocities. The viscosity coefficients in Eqs. (2) and (3), in general, include contributions from both the intra- and inter-species collisions. However, provided $Z \gg 1$ and assuming comparable hydrogen and high-Z species number densities, we neglect small contributions from weak H-H (relative to H-Z) and Z-H (relative to Z-Z) collisions, and estimate the collisional viscosity coefficients as $\eta_H \approx \eta_{HZ} \sim M_H V_{TH} n_H \lambda_{HZ}$ and $\eta_Z \approx \eta_{ZZ} \sim M_Z V_{TZ} n_Z \lambda_{ZZ}$.

From Eqs. (2)-(3) it follows that slow relaxation of a shear in the common flow velocity occurs on a transport time-scale, $\tau_{shear} \sim (M_Z n_Z/M_H n_H)(L/\lambda_{HZ})^2 \tau_c \gg \tau_c$ and is governed by

$$\partial[(M_H n_H + M_Z n_Z)V_x]/\partial t = \frac{\partial}{\partial y}\eta\frac{\partial V_x}{\partial y} \sim \frac{\partial}{\partial y}(M_Z V_{TZ} n_Z \lambda_{ZZ} + M_H V_{TH} n_H \lambda_{HZ})\frac{\partial V_x}{\partial y}. \qquad (4)$$

It is also straightforward to show that the small difference between the hydrogen and high-Z species flow velocities is the order of $\Delta V_x/V_x \sim \lambda_{HZ}^2/L^2$. Assuming equal temperatures of the hydrogen and high-Z ions, we readily obtain $\lambda_{HZ} \sim Z^2 \lambda_{ZZ}$, $V_{TH} \sim \sqrt{M_Z/M_H} V_{TZ}$, and

$$\frac{\eta_{HZ}}{\eta_{ZZ}} \sim \frac{n_H}{n_Z} Z^2 \sqrt{\frac{M_H}{M_Z}}. \tag{5}$$

For instance, for a CD$_2$ plasma ($n_H/n_Z = 2$, $M_H/M_Z = 2/12$, $Z = 6$), it follows that $\eta_{HZ}/\eta_{ZZ} = 30$ [9]. That is the hydrogen species provides the dominant contribution to the plasma viscosity,

$$\eta \approx \eta_{HZ}. \tag{6}$$

Note that the energy exchange between species occurs much faster, $\tau_E \sim (M_Z/M_H)\tau_c$, than relaxation of the velocity shear, $\tau_c \ll \tau_{shear}$, and therefore the assumption of equal temperatures, $T_Z = T_H$, used in deriving Eq. (5) is justified.

While an accurate expression for the shear viscosity coefficient in a single-ion-species plasma is well-known [10],

$$\eta_{ZZ} = 0.96 \frac{3}{4\pi^{1/2}} \frac{\sqrt{M_Z} T_Z^{5/2}}{Z^4 e^4 \ln \Lambda_Z}, \tag{7}$$

where $\ln\Lambda_Z$ denotes the Coulomb logarithm, the goal of the present work is to derive an equivalent accurate expression for the viscous coefficient corresponding to the H-Z collisions, i.e., $\eta_{HZ}$.

*Quantitative analysis.*–It is well-known that the viscosity of a monatomic gas or unmagnetized plasma is described by a single shear-viscosity coefficient, $\eta$, with the viscous stress tensor given by $\sigma_{ij} = -\eta(\partial V_i/\partial x_j + \partial V_j/\partial x_i - 2/3\delta_{ij}\nabla \cdot \mathbf{V})$ [10-11]. From the standard framework for determining transport coefficients [10-12], it follows that in order to evaluate the viscosity coefficient, $\eta_{HZ}$, one needs to find a first-order correction in $\lambda_{HZ}/L \ll 1$, i.e., $F_1$, to a zero-order solution of the kinetic equation given by a local Maxwellian distribution with a shear flow velocity

$$F_M(\mathbf{r},\mathbf{u}) = \frac{n_H}{\pi^{3/2} V_{TH}^3} \exp\left(-\frac{[u_x - V_x(y)]^2 + u_y^2 + u_z^2}{V_{TH}^2}\right). \tag{8}$$

Here, $\mathbf{r}$ and $\mathbf{u}$ are the particle's coordinate and velocity, $V_{TH} = \sqrt{2T_H/M_H}$ is the hydrogen thermal velocity, and, without loss of generality, we adopted $\mathbf{V} = V_x(y)\hat{\mathbf{x}}$. Introducing $\bar{u}_x = u_x - V_x(y)$, one obtains the following equation for the first-order correction $F_1$

$$2\frac{\bar{u}_x u_y}{V_{TH}^2}\frac{\partial V_x}{\partial y}F_M = \nu_{HZ}C[F_1], \tag{9}$$

where, $\nu_{HZ}(u) = (2\pi n_Z Z^2 e^4 / M_H^2 u^3)\ln\Lambda_H$, and $C[F]$ denotes a normalized collision operator for H-Z collisions. In what follows we consider a succession of increasingly detailed collision operator options. First, we assume a small mass ratio $M_H/M_Z \ll 1$, and adopt the Lorentz collision operator to obtain an exact solution to Eq. (9). A first-order correction in $M_H/M_Z \ll 1$ is then analytically recovered by assuming a Maxwellain distribution of the high-Z ions and considering hydrogen collisions with this Maxwellian background. Finally, the same collision model is used to obtain numerical solutions to Eq. (9) for the case of an arbitrary mass ratio.

*Case (a): Lorentz model.*–Assuming, for simplicity, $M_H \ll M_Z$ we adopt the Lorentz collision model,

$$C[F_1] = \frac{1}{\sin\theta}\frac{\partial}{\partial\theta}\sin\theta\frac{\partial F_1}{\partial\theta} + \frac{1}{\sin^2\theta}\frac{\partial^2 F_1}{\partial\varphi^2}, \tag{10}$$

where, $(u,\theta,\varphi)$ are the spherical coordinates corresponding to the coordinate system $(\bar{u}_x, u_y, u_z)$. In general, the collision operator in Eq. (10) should also include terms proportional to the relative velocity between the H and Z species (see Refs. [10]-[12]). However, these terms are of the second order in $\lambda_{HZ}/L$ (recall that $\Delta V_x/V_x \sim \lambda_{HZ}^2/L^2$), and therefore are neglected in Eq. (10). Noting that

$$\bar{u}_x u_y/u^2 = \cos\varphi\sin\theta\cos\theta = \sqrt{\frac{8\pi}{15}}\frac{Y_{2,-1} - Y_{2,1}}{2}, \tag{11}$$

where $Y_{l,m}(\theta,\varphi)$ are the spherical harmonic functions [14] that satisfy $C(Y_{l,m}) = -l(l+1)Y_{l,m}$, it follows that

$$F_1 = -\sqrt{\frac{8\pi}{15}}\frac{Y_{2,-1} - Y_{2,1}}{6}\frac{u^2}{V_{TH}^2}\frac{F_M}{\nu_{HZ}(u)}\frac{\partial V_x}{\partial y}. \tag{12}$$

The viscous momentum flux is given by

$$\sigma_{xy} = \sigma_{yx} = M_H\int\bar{u}_x u_y F_1 d^3\mathbf{u} = -\eta_{HZ}\frac{\partial V_x}{\partial y}, \tag{13}$$

and we readily obtain

$$\eta_{HZ} = M_H\oint d\Omega(Y_{2,-1} - Y_{2,1})^2\int_0^\infty du\,\frac{2}{45}\pi\frac{u^6}{V_{TH}^2}\frac{F_M(u)}{\nu_{HZ}(u)}, \tag{14}$$

where $d\Omega = \sin\theta d\theta d\varphi$. Making use of the properties of spherical harmonics, i.e., $Y_{l,-m} = (-1)^m Y_{l,m}^*$ and $\oint d\Omega Y_{l',m'}^* Y_{l,m} = \delta_{l,l'}\delta_{m,m'}$, it follows that $\oint d\Omega (Y_{2,1} - Y_{2,-1})^2 = 2$. Performing the remaining integration along the u-axis, we obtain

$$\eta_{HZ} = \frac{32\sqrt{2}}{15\pi^{3/2}} \frac{n_H}{n_Z} \frac{\sqrt{M_H} T_H^{5/2}}{Z^2 e^4 \ln\Lambda_H}, \tag{15}$$

or, equivalently,

$$\eta_{HZ}(s^{-1}cm^{-1}g) = \frac{4.27\times 10^{-5}}{\ln\Lambda_H} \frac{n_H}{n_Z} \frac{A_H^{1/2}}{Z^2} (T_H[eV])^{5/2}, \tag{16}$$

where $A_H$ is the atomic mass of the hydrogen isotope. Recalling Braginski's result for $\eta_{ZZ}$ [Eq. (7)], a more accurate form of the estimate in Eq. (5) is given by

$$\frac{\eta_{HZ}}{\eta_{ZZ}} = 1.33 \frac{n_H}{n_Z} Z^2 \sqrt{\frac{M_H}{M_Z}}, \tag{17}$$

where $\ln\Lambda_Z = \ln\Lambda_H$ is assumed for simplicity. It is also straightforward to extend the result in Eq. (15) for the case where several high-Z ion species are present

$$\eta_{HZ} = \frac{32\sqrt{2}}{15\pi^{3/2}} \frac{n_H}{\sum_j Z_j^2 n_{Zj}} \frac{\sqrt{M_H} T_H^{5/2}}{e^4 \ln\Lambda_H}. \tag{18}$$

*Case (b): Collisions with a Maxwellian background, $M_H/M_Z \ll 1$.*–The expressions for the viscosity coefficient in Eqs. (15)-(18) are only valid in the limit $M_H \ll M_Z$, and do not include any effects associated with a finite mass ratio. A first-order correction that includes the effects appearing in order $O(M_H/M_Z)$ can be readily derived as follows. Assuming that the charge state of the Z-species ions is sufficiently high, and thus Z-Z collisions are sufficiently strong, we neglect deviations of the high-Z species distribution from a Maxwellian. The collision operator in Eq. (10) describing H-Z collisions can then be generalized to include effects of an arbitrary mass ratio as follows [12]

$$C[F_1] = \nu_D \left[\frac{1}{\sin\theta}\frac{\partial}{\partial\theta}\sin\theta\frac{\partial F_1}{\partial\theta} + \frac{1}{\sin^2\theta}\frac{\partial^2 F_1}{\partial\varphi^2}\right] + u\frac{\partial}{\partial u}\left[u^3\left(\frac{M_H}{M_H+M_Z}\nu_s F_1 + \frac{1}{2}\nu_\parallel u \frac{\partial F_1}{\partial u}\right)\right]. \tag{19}$$

Here, $\nu_D = Y(\xi) - G(\xi)$, $\nu_s = 2[(M_H+M_Z)/T_Z]G(\xi)/u$ and $\nu_\parallel = 4G(\xi)/u^3$ are the collision coefficients, $\xi = u/\sqrt{2T_Z/M_Z}$ is the normalized particle speed, and the dimensionless functions $Y(\xi)$ and $G(\xi)$ are specified by $Y(\xi) = (2/\sqrt{\pi})\int_0^\xi e^{-t^2} dt$ and $G(\xi) = [Y(\xi) - xY'(\xi)]/2\xi^2$.

Although the collision model in Eq. (19) is valid for an arbitrary value of $M_H/M_Z$, to progress analytically here we assume $M_H/M_Z \ll 1$. The operator in Eq. (19) then takes on the following form:

$$C[F_1] = \left(1 - \frac{V_{TH}^2}{2u^2}\frac{M_H}{M_Z}\right)\left(\frac{1}{\sin\theta}\frac{\partial}{\partial\theta}\sin\theta\frac{\partial F_1}{\partial\theta} + \frac{1}{\sin^2\theta}\frac{\partial^2 F_1}{\partial\varphi^2}\right) + 2\frac{M_H}{M_Z}u\frac{\partial}{\partial u}\left[F_1 + \frac{T_Z}{M_H}\frac{1}{u}\frac{\partial F_1}{\partial u}\right]. \quad (20)$$

Representing the solution to Eq. (9) as $F_1 = F_1^0 + (M_H/M_Z)F_1^{(1)}$, where $F_1^0$ is given by Eq. (12), the correction $F_1^{(1)}$ is determined from

$$\left(\frac{1}{\sin\theta}\frac{\partial}{\partial\theta}\sin\theta\frac{\partial}{\partial\theta} + \frac{1}{\sin^2\theta}\frac{\partial^2}{\partial\varphi^2}\right)\left(F_1^{(1)} - \frac{V_{TH}^2}{2u^2}\frac{M_H}{M_Z}F_1^0\right) + 2u\frac{\partial}{\partial u}\left[F_1^0 + \frac{T_Z}{M_H}\frac{1}{u}\frac{\partial F_1^0}{\partial u}\right] = 0. \quad (21)$$

Recalling that equilibration of species temperatures occurs much faster than relaxation of a common velocity shear, we take $T_Z = T_H = T$, and after some straightforward algebra obtain

$$\eta_{HZ} = \frac{32\sqrt{2}}{15\pi^{3/2}}\frac{n_H}{n_Z}\frac{\sqrt{M_H}T^{5/2}}{Z^2 e^4 \ln\Lambda_H}\left(1 - \frac{11}{12}\frac{M_H}{M_Z}\right). \quad (22)$$

In the case several high-Z ion species are present, the viscosity in Eq. (22) takes the following form

$$\eta_{HZ} = \frac{32\sqrt{2}}{15\pi^{3/2}}\frac{n_H}{\sum_j Z_j^2 n_{Zj}}\frac{\sqrt{M_H}T^{5/2}}{e^4 \ln\Lambda_H}\left(1 - \frac{11}{12}\sum_j \frac{M_H}{M_{Zj}}Z_j^2 n_{Zj} \bigg/ \sum_j Z_j^2 n_{Zj}\right). \quad (23)$$

*Case (c): Collisions with a Maxwellian background, $M_H/M_Z \sim 1$.* – The expressions in Eqs. (22)-(23) omit higher-order corrections in $M_H/M_Z$ [i.e., $O(M_H^2/M_Z^2)$, $O(M_H^3/M_Z^3)$, ...]. However, these terms may be important for a detailed analysis of the lithium-hydride HED plasmas that will appear in an ICF reactor if a lithium-hydride blanket is used [4-5]. In order to address this practical issue, which is also of general theoretical interest, we now evaluate accurate values of the H-Z viscosity coefficient for an arbitrary $M_H/M_Z$ mass ratio, by making use of the variational principle [13]. Introducing $\bar{f}_1 = \bar{F}_1/F_M$, where $\bar{F}_1$ is an arbitrary trial function, and noting that the collision operator in Eq. (19) is self adjoint with

$$\int d^3\mathbf{u} \bar{f}_1 \nu_{HZ}(u) C[\bar{F}_1] = \int d^3\mathbf{u} \bar{F}_1 \nu_{HZ}(u) C[\bar{f}_1], \quad (24)$$

it follows that Eq. (9) is the Euler equation for the variational principle

$$\delta S \equiv \delta(S_1 - 2S_2) = 0, \tag{25}$$

where the functionals $S_1$ and $S_2$ are defined by

$$S_1 = \int d^3\mathbf{u}\bar{f}_1 \nu_{HZ}(u) C[\bar{F}_1], \tag{26}$$

$$S_2 = \int d^3\mathbf{u}\bar{f}_1 \frac{2\bar{u}_x u_z}{V_{TH}^2} \frac{\partial V_x}{\partial y} F_M. \tag{27}$$

It is important to note that the extremal value of the variational quantity, $S^{ext}$, satisfies

$$S^{ext} = -S_2^{ext} = \left(\frac{2}{M_H V_{TH}^2}\frac{\partial V_x}{\partial y}\right)\eta, \tag{28}$$

where $S^{ext}$ and $S_2^{ext}$ corresponds to the functionals in Eq. (25) and Eq. (27) evaluated with the exact solution to Eq. (9). The fact that the viscosity coefficient is directly related to the extremal value of the variational quantity provides very good accuracy of this asymptotic method.

We now proceed with the variational analysis by making use of the following trial function $\bar{F}_1$ [10]

$$\bar{F}_1(\mathbf{u}) = [Y_{2,-1}(\theta,\varphi) - Y_{2,1}(\theta,\varphi)] u^2 \sum_{i=1}^{N} a_i L_i^{(5/2)}(u^2/V_{TH}^2), \tag{29}$$

where $L_i^{(5/2)}(x)$ are Sonine polynomials, $L_0^{(5/2)}(x) = 1$, $L_1^{(5/2)}(x) = 7/2 - x$, $L_2^{(5/2)}(x) = 63/8 - 9x/2 + x^2/2$, $N$ determines the number of terms retained the expansion, and the coefficients $a_i$ are found from a linear system of equations, $\partial S/\partial a_i = 0$. Due to orthogonally of the Sonine polynomials, the viscosity coefficient, $\eta_{HZ}$, is determined only by the value of $a_0$. While an exact value of $\eta_{HZ}$ would be recovered in the limit $N \to \infty$, it is found that sufficient accuracy can be obtained by retaining only a few first terms in the expansion in Eq. (29). The results for the viscosity coefficient can be expressed as

$$\eta_{HZ} = \frac{32\sqrt{2}}{15\pi^{3/2}} \frac{n_H}{n_Z} \frac{\sqrt{M_H} T^{5/2}}{Z^2 e^4 \ln\Lambda_H}(1-B), \tag{30}$$

where the value of the coefficient $B$ is given in Table 1 for various choices of $M_H/M_Z$. It is important to emphasize here that the result in Eq. (30) was obtained by making use of the collision operator in Eq. (19), i.e., neglecting deviations of the high-Z species distribution from a Maxwellian. Also, equal temperatures of hydrogen and high-Z species were assumed.

Table 1. Numerical values of the coefficient $B$ [in Eq. (30)] obtained with $N=4$.

| $M_Z/M_H$ | 1 | 2 | 3 | 4 | 5 | 6 | 7 | 8 | 9 | >>1 |
|---|---|---|---|---|---|---|---|---|---|---|
| $B$ | 0.374 | 0.283 | 0.220 | 0.178 | 0.150 | 0.129 | 0.113 | 0.101 | 0.091 | $\frac{11}{12}\frac{M_H}{M_Z}$ |

Note that unlike the results in Eqs. (15) and (22), the result in Eq. (30) cannot be extended straightforwardly to the more general case of several high-Z ion species. Each new combination of high-Z species requires separate use of the variational principle to determine the corresponding viscosity coefficient. However, we emphasize that the first-order accurate estimate of the viscosity coefficient for the case where several heavy high-Z ion species with $M_H/M_{Zj} \ll 1$ are present is given in Eq. (23), or equivalently

$$\eta_{HZ}(s^{-1}cm^{-1}g) = \frac{4.27 \times 10^{-5}}{\ln \Lambda_H} \frac{n_H A_H^{1/2}(T[eV])^{5/2}}{\sum_j Z_j^2 n_{Zj}} \left(1 - \frac{11}{12}\sum_j \frac{M_H}{M_{Zj}} Z_j^2 n_{Zj} \Big/ \sum_j Z_j^2 n_{Zj}\right), \quad (31)$$

where $A_H$ is the atomic mass the hydrogen isotope.

In conclusion, as an illustrative example, we present numerical values of the viscosity coefficients for materials that are often considered for HED and ICF applications (see Table 2).

Table 2. Viscosity coefficients for several choices of a hydrogen-containing material[a],

$$\eta_{HZ}(s^{-1}cm^{-1}g) = \frac{\alpha \times 10^{-6}}{\ln \Lambda_H}(T[eV])^{5/2}$$

| Material | CH | $CH_2$ | CD | $CD_2$ | LiH | $C_{22}H_{10}N_2O_4$ |
|---|---|---|---|---|---|---|
| $\alpha$ | 1.10 | 2.20 | 1.56 | 3.12 | 4.20 | 0.346 |

[a]The coefficient $\alpha$ is obtained by using Eq. (31) for the case of polyimide ($C_{22}H_{10}N_2O_4$), and by using the variational method for all other materials.

**Acknowledgment.** The author appreciates helpful discussions with Dr. D. Ryutov. This work is performed for the U.S. DOE at LLNL under contract DE-AC52-07NA27344.